\documentclass[prb,twocolumn,superscriptaddress, showpacs]{revtex4}

\usepackage{epstopdf}
\usepackage{verbatim}

\usepackage{amsmath}
\usepackage{amsfonts}

\usepackage{graphicx}
\usepackage{times}

\usepackage{units}

\mathchardef\mhyphen="2D

\newcommand{\be}{\begin{equation} }
\newcommand{\ee}{\end{equation} }

\newcommand{\ba}{\begin{eqnarray} }
\newcommand{\ea}{\end{eqnarray} }

\begin{document}

\title{Quantum Cognition: The possibility of processing with nuclear spins in the brain 
}

\author{Matthew P.\ A.\ Fisher}
\affiliation{Department of Physics, University of California, Santa Barbara, CA 93106}

\date{\today}
\begin{abstract}
The possibility that quantum processing with nuclear spins might be operative in the brain is explored.  Phosphorus is identified as the unique biological element with a nuclear spin that can serve as a qubit for such putative quantum processing - {\it a neural qubit} -  while
the phosphate ion is the only possible {\it qubit-transporter}.  We identify the
``Posner molecule", $\text{Ca}_9 (\text{PO}_4)_6$, as the unique molecule that can protect the neural qubits on very long times and thereby serve as a (working) {\it quantum-memory}.
A central requirement for quantum-processing
is {\it quantum entanglement}.  It is argued that the 
enzyme catalyzed chemical reaction which breaks a pyrophosphate ion into two phosphate ions
can quantum entangle pairs of qubits.   Posner molecules, formed by binding such
phosphate pairs with extracellular calcium ions,
will inherit the nuclear spin entanglement.
A mechanism for transporting Posner molecules into 
presynaptic neurons 
during vesicle endocytosis is proposed.
{\it Quantum measurements} can occur when a pair of Posner molecules chemically bind and subsequently melt, releasing a shower of intra-cellular calcium ions that can trigger further neurotransmitter
release and enhance the probability of  post-synaptic neuron
firing.   Multiple entangled Posner molecules, triggering non-local quantum correlations of neuron firing rates, would provide the key mechanism for neural quantum processing.
Implications, both in vitro and in vivo, are briefly mentioned.

\end{abstract}

\maketitle
 
%\tableofcontents
{\it ``I can calculate the motion of heavenly bodies, but not the madness of men" - Isaac Newton}

\subsection{Introduction}

\label{sec:intro}

It has long been presumed that quantum mechanics cannot play an important (functional) role
in the brain, since maintaining quantum coherence on macroscopic time scales (seconds, minutes, hours,...) is exceedingly unlikely in a wet environment\cite{Qu_Brain_2000,Qu_Brain_2000a} (although see [\onlinecite{Hu_Wu_2004, Penrose_2011}] and references therein).
Small molecules, or even individual ions, while described in principle by quantum mechanics, rapidly entangle with the surrounding environment,
which causes de-phasing of any putative quantum coherent phenomena.
However, there is one exception: Nuclear spins are so weakly coupled to the
environmental degrees of freedom that, under some circumstances, phase coherence
times of five minutes or perhaps longer are possible.\cite{NMR_Hore,NMR_Li_1976}

Putative quantum processing with nuclear spins in the wet environment of the brain - as proposed by Hu and Wu in Ref. [\onlinecite{Hu_Wu_2004}] - would seemingly require fulfillment of many unrealizable conditions: for example, a common biological element with a long nuclear-spin coherence time to serve as a qubit, a mechanism for transporting this qubit throughout the brain and into neurons, a molecular scale quantum memory for storing the qubits, a mechanism for quantum entangling multiple qubits, a chemical reaction that induces quantum measurements on the qubits which 
dictates subsequent neuron firing rates, among others.

Our strategy, guided by these requirements and detailed below, is one of ``reverse engineering" - 
seeking to identify the bio-chemical ``substrate" and mechanisms hosting such putative
quantum processing.  Remarkably, a specific neural qubit and a unique collection
of ions, molecules, enzymes and neurotransmitters is identified,
illuminating an apparently single path
towards nuclear spin quantum processing in the brain.

\subsection{The Neural qubit - phosphorus nuclear-spin}

The nucleus of every element is
characterized by a half-integer spin-magnitude ($\text{I}=0,\frac{1}{2},1\cdots$)
and for $\text{I} \ne 0$ an associated magnetic dipole moment which precesses around magnetic fields at the nucleus.\cite{QM_Gottfried}  These magnetic fields arise from nuclear magnetic moments of nearby atoms/ions.
Nuclei with $\text{I} > \frac{1}{2}$ also have an electric quadrupole moment which couples to electric field gradients generated by charges of nearby electrons/nuclei.\cite{NMR_Hore}
Magnetic and electric field perturbations 
cause quantum decoherence of the nuclear spin - anathema to quantum processing - 
so that the  ``coherence time", $t_{\text{coh}}$, must be maximized when seeking a possible biological arena for nuclear spin processing.

In the biochemical setting 
electric fields are the primary source of decoherence for nuclei with $\text{I} >\frac{1}{2}$, while $\text{I}=\frac{1}{2}$ spins are more weakly decohered only by magnetic fields.
For example, a solvated $^7 \text{Li}^+$ isotope with $\text{I}=\frac{3}{2}$ has $t_\text{coh} \sim 10 \text{s}$, while the $^6 \text{Li}^+$ isotope (with very small electric quadrupole moment) is an ``honorary $\text{I}=\frac{1}{2}$" with $t_\text{coh}$ as long as 5 minutes!\cite{NMR_Li_1976}
Thus, {\it the element hosting the putative neural qubit must have nuclear-spin} $\text{I}=\frac{1}{2}$.

Among the most common biochemical elements, carbon, hydrogen, nitrogen, oxygen, phosphorus and sulphur, and the common ions $\text{Na}^+,\text{K}^+,\text{Cl}^{-},\text{Mg}^{2+}$ and $\text{Ca}^{2+}$, besides hydrogen, only phosphorus has a nucleus with spin $\text{I}=\frac{1}{2}$.  {\it This  identifies the phosphorus nucleus as our putative neural qubit}.

\subsection{Qubit transporter - the phosphate ion}

Phosphorus is bound into the inorganic phosphate
ion $\text{PO}_4^{3-}$ (abbreviated as Pi) in biochemistry, present in energy transport molecules such as ATP and in poly-phosphate chains
including the pyrophosphate ion, $\text{P}_2\text{O}_7^{4-}$
(abbreviated as PPi).\cite{Alberts_Cell,Lodish_Cell}
The tetrahedrally coordinated oxygen cage surrounding phosphorus in Pi
resembles the cage of oxygens in the hydration shell of solvated $^6 \text{Li}^+$.
However, the phosphorus spin coherence time for solvated Pi, $t_\text{coh}  \sim 1 \text{s}$,  
is significantly shorter
than the coherence time of solvated $^6 \text{Li}^+$, $\sim 5 \text{min}$.  This difference can be attributed
to the proton that binds to Pi at physiological pH -  the proton and phosphorus nuclear spins in $\text{HPO}_4^{2-}$ ($\text{HPi}$) interact via the electrons and
significantly reduce the phosphorus spin coherence time
to $\sim 1 \text{s}$.

The solvated phosphate ion can nevertheless serve as an effective qubit transporter diffusing $\sim 10 \mu \text{m}$ (the cellular scale) in roughly $10^{-2} \text{s}$,\cite{extracellular_space_1998}  while maintaining spin coherence.   Qubit memory-storage (and processing) on times of seconds or longer will, however, require another molecule,
which we next discuss.

\subsection{Qubit memory - the Posner molecule}

If another biological cation (with $\text{I}=0$) can displace the proton in binding
to the phosphate ions, longer spin coherence times might be possible.
The presence of bone mineral, calcium-phosphate,\cite{Bone_mineral_2003}  
indicates that under some physiological conditions calcium ions can out-compete the proton in binding to phosphates.
Indeed, several recent in vitro studies have found evidence for a stable calcium-phosphate molecule,\cite{Posner_expt_1998,Posner_expt_1999,Posner_expt_2010,Posner_expt_2012}  $\text{Ca}_9 (\text{PO}_4)_6$ (see Figure 1),  in simulated body fluids (SBF) appropriate for the extracellular fluid of mammals.
These nanometer diameter ``Posner molecules"
are likely present in real extracellular body fluid as well.

The phosphorus spins in a Posner molecule are expected to have
{\it very long} coherence times, which we estimate as follows.
The magnetic dipole fields from protons in nearby water molecules will
cause the phosphorus spins to precess at frequencies of order $f_\text{dip} \sim 10^3 \text{Hz}$, naively suggesting milli-second coherence times.
But due to the rapid tumbling of Posner molecules in water (with rotation frequencies of order
$f_\text{rot} \sim 10^{11} \text{Hz}$), the magnitude and direction of the dipole magnetic field at a given phosphorus nucleus will vary rapidly with time ($f_\text{rot}^{-1} \sim 10 \text{psec}$), averaging to zero.
Residual magnetic field fluctuations will lead to 
 ``directional diffusion" of the phosphorus spins with very long coherence times,
$t_\text{\text{coh}} \simeq f_\text{rot}/ f^2_\text{dip}  \sim   10^5 s \sim 1 \text{day}$.
Actual coherence times could well be even longer, since 5 of the 64 spin states
in the Posner molecule, with zero (total) spin, $\text{I}_\text{tot} = 0$, 
will be virtually blind to decoherence.

\begin{figure}
\includegraphics[width=3.5in]{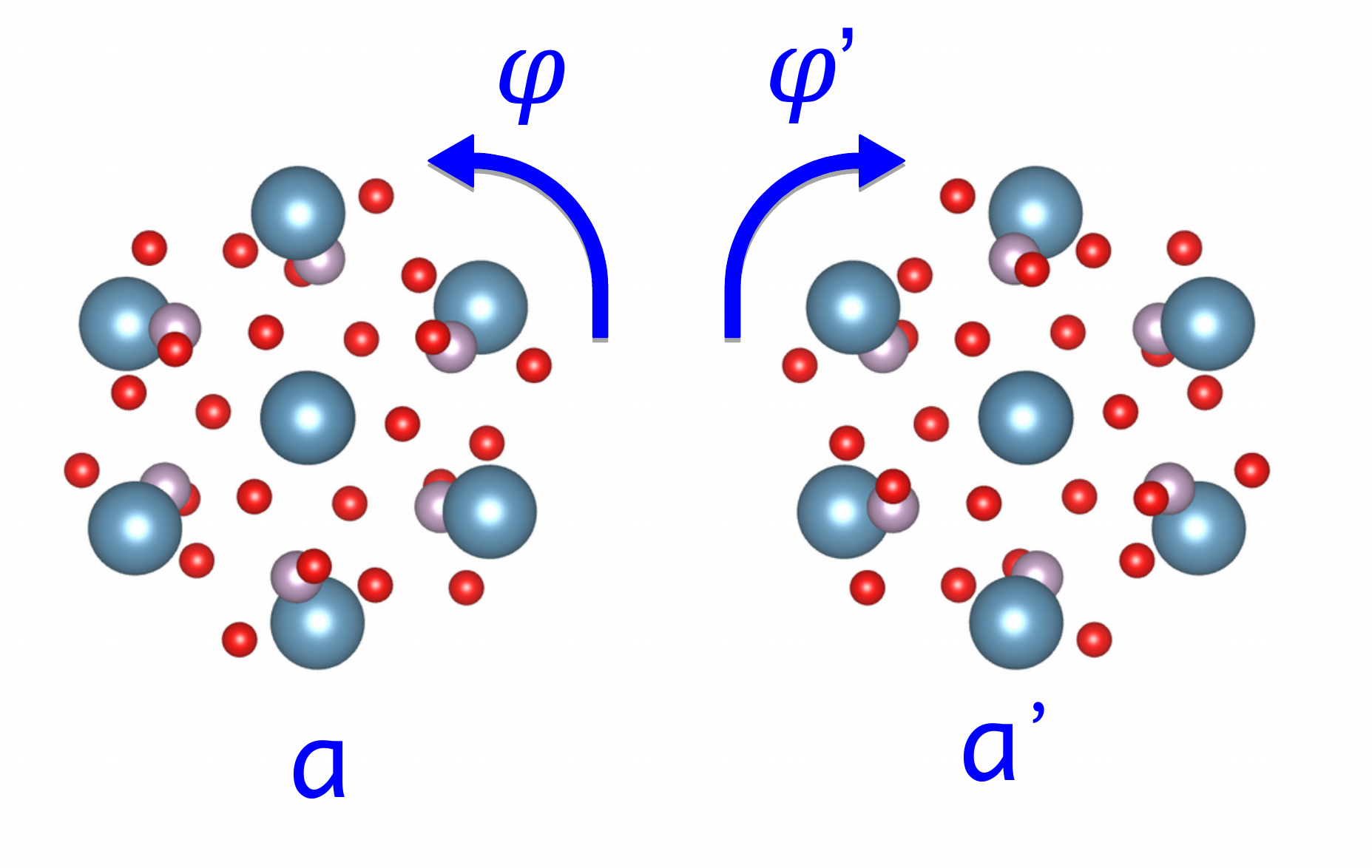}
\caption{
Two Posner molecules, $\text{Ca}_9 (\text{PO}_4)_6$, with calcium ions (blue) in a bcc arrangement
(eight at the corners and one at the center of a cube) as viewed along the (111) axis.
The six phosphate ions   -  a tetrahedron of oxygens (red) surrounding a central phosphorus (purple) - 
are on the cube faces and reduce the symmetry to $\text{S}_6$, with one remaining 3-fold symmetry axis.
As shown, the two Posner molecules are oriented with the 3-fold axis out of the page (molecule $a$) 
and into the page (molecule $a^\prime$), with
$\varphi$ and $\varphi^\prime$ denoting the rotation angles around their respective symmetry axis.
In this orientation quantum chemistry calculations indicate that the two Posner molecules can
chemically bind to one another, releasing of order an $eV$ of energy.\cite{M_Swift}
} 
\label{Posner_pair}
\end{figure}

\subsection{Enzyme catalyzed qubit entanglement}

 ``Quantum entanglement" between qubits is necessary for quantum processing.\cite{Qu_Computing, RMP_Qu_Entanglement_2007}
While a single qubit state can be expressed in terms 
of  ``up"  and ``down" basis states,
a pair of qubits will have four basis states, $|\uparrow \uparrow \rangle , | \uparrow \downarrow \rangle$ and so on.
The two-qubit ``spin-singlet" state,
$| s \rangle = [ |\uparrow \downarrow \rangle - |\downarrow \uparrow \rangle ]/\sqrt{2}$,
embodies a form of entanglement which
lies at the heart of quantum mechanics and
serves as the ``unit of currency" for laboratory quantum computing efforts.
If the first spin is measured to be ``up", 
then the second spin will be found ``down"  - and vice versa -  independent of the spatial separation of the two spins - a  non-local entanglement referred to by Einstein as 
 ``spooky action at a distance".\cite{QM_Gottfried}
 
Two-qubit states describe the phosphorus nuclear spins in the important biochemical ion
pyrophosphate (PPi), a linear phosphate-dimer created in the hydrolysis reaction $\text{ATP} \rightarrow \text{AMP} + \text{PPi}$.\cite{Alberts_Cell,Lodish_Cell}
The four time-independent (stationary) states of the two interacting phosphorus spins are the (para) spin-singlet and three (ortho) spin-triplet states, $|t_+\rangle = | \uparrow \uparrow \rangle$,  $|t_-\rangle = | \downarrow \downarrow \rangle$ and $|t_0 \rangle =  [ | \uparrow \downarrow \rangle + | \downarrow \uparrow \rangle ]/\sqrt{2}$.  A general spin state of PPi can be written as a linear combination of these four
stationary states.

Quantum mechanics is usually presumed irrelevant
in describing the translational, vibrational and 
rotational motion of molecules or ions diffusing in water.
However, an adequate description of the hydrolysis reaction of interest to us,\cite{Alberts_Cell,Lodish_Cell,Pyrophosphatase_1966}
$\text{PPi} \rightarrow \text{Pi} + \text{Pi}$, {\it requires 
a full quantum treatment of the molecular rotations}.
With rotations included the quantum state of PPi can be presented as a sum of singlet and triplet terms of the form,
\begin{equation}
 \Psi_\text{PPi}  = c_s \psi_s (\hat{\bf r}) |s \rangle   + c_t \psi_t  (\hat{\bf r}) | t \rangle  ,
\label{Psi_PPi}
\end{equation}
where $| t \rangle = \sum_m a_m |t_m \rangle$  ($m=0,\pm1$) with normalizations,
$\langle t | t \rangle =1$ and $|c_s|^2 + |c_t|^2=1$. 
Here $\hat{\bf r}$ is a unit vector 
specifying the ions orientation.  Being identical fermions,
$\Psi_{\text{PPi}}$ must change sign under the interchange of the two phosphorus atoms - which corresponds to an exchange of the spins and a 180 degree body rotation, $\hat{\bf r} \rightarrow - \hat{\bf r}$.
Since the spin-singlet changes sign under exchange,
$| s \rangle \rightarrow -  | s \rangle$, while the spin-triplets do not, the corresponding orbital wavefunctions must satisfy, $\psi_{s/t} (\hat{\bf r}) = \pm \psi_{s/t} (-\hat{\bf r})$, so that both terms in Eq. (\ref{Psi_PPi}) change sign under full exchange, as required.\cite{E_Bright_Wilson}
The first and second terms in $\Psi_\text{PPi}$ are direct analogs of the para and ortho states
of molecular hydrogen - perhaps we might call them para and ortho pyrophosphate -  but,
quite generally, the appropriate state of both $\text{H}_2$ and $\text{PPi}$ should be a quantum linear superposition of the para and ortho states.

The stationary states of freely rotating PPi  (``spherical harmonics")
are labelled by an integer angular momentum $L \ge 0$ (in units of $\hbar$).
Under a 180 degree rotation the 
spherical harmonics are multiplied by $(-1)^L$, so that
$\psi_s$ and $\psi_t$ can be expressed in terms of even 
and odd angular momentum wavefunctions, respectively.
For PPi tumbling in water, $L \sim  100$,
so the distinction between even and odd values of $L$ is unimportant.

However, the enzyme catalyzed hydrolysis reaction $\text{PPi} \rightarrow \text{Pi} + \text{Pi}$ 
to which we next turn, requires first slowing ($L \sim O(1)$) and then stopping the PPi rotation,
a process which will presumably depend on the sign of $(-1)^L$.
This suggests {\it a reaction rate dependent
on the nuclear spin state, different for the singlet and triplet states}.

\subsubsection{The enzyme pyrophosphatase}

We illustrate this within a simple model of the enzyme pyrophosphatase.\cite{Pyrophosphatase_2009}
The four magnesium ions inside the enzyme pocket,
each with charge $+1$ 
appropriate when singly bonded to an enzyme oxygen, 
will attract pyrophosphate $\text{P}_2 \text{O}_7^{4-}$ into the pocket (see Figure 2).\cite{Pyrophosphatase_2009}
When rapidly rotating PPi will 
 ``look" like
a spherical shell of charge,
held in place by the magnesium ions, symmetrically arranged for maximum stability, as depicted in Figure 2a.
But in the aspherical electrostatic environment in the pocket,  
PPi will lose angular momentum and, once slowed, will tend to align along preferred orientations, held by the ``pinning" potentials attracting the negatively charged PPi oxygens to the positive magnesium ions, of strength $v_x$ and $v_y$ (see Figure 2b and 2c).

\begin{figure}
\includegraphics[width=3.5in]{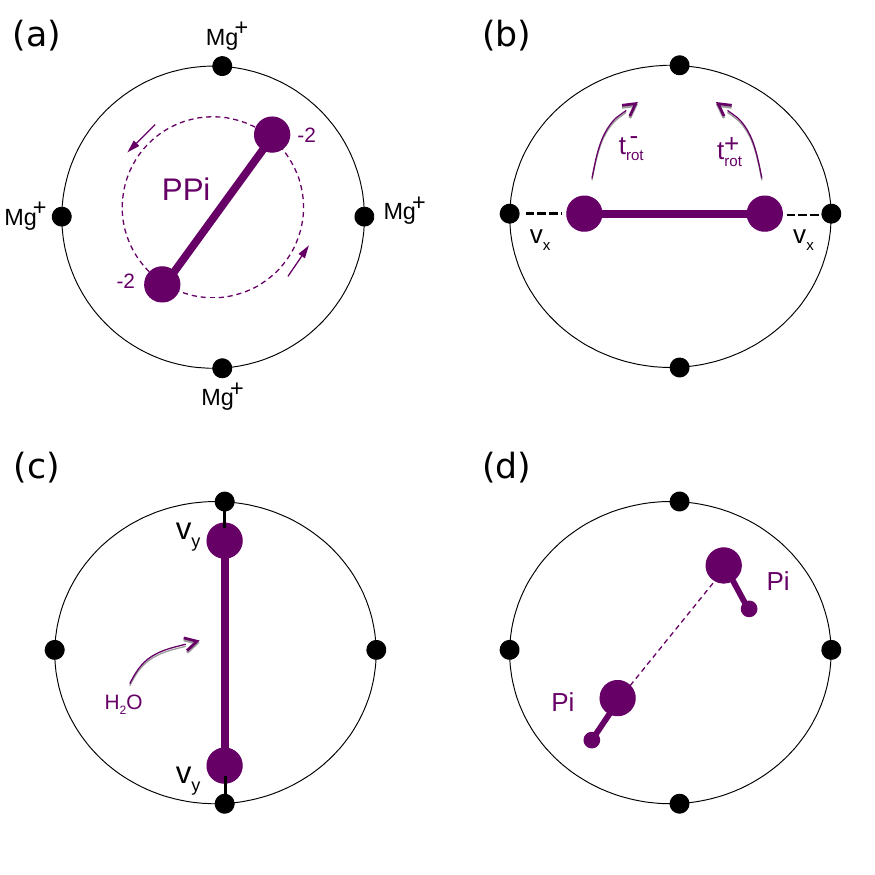}
\caption{
The pyrophosphate ion $\text{P}_2 \text{O}_7^{4-}$ (PPi), shown rotating inside the pocket of the enzyme pyrophosphatase in (a), is attracted  
by four enzyme bound $\text{Mg}^+$ ions.    Once the rotation slows, PPi
will orient along preferred directions to align with the $\text{Mg}^+$ ions, as in (b) and (c), with $v_x, v_y$ denoting respective ``pinning" strengths.  When PPi binds (chelation) to two 
$\text{Mg}^+$ ions, as depicted in (c), the weakened internal covalent bonds of PPi will
facilitate the hydrolysis reaction, 
$\text{PPi} + \text{H}_2\text{O} \rightarrow \text{Pi} + \text{Pi}$.
The phosphorus nuclear spins in the two liberated phosphate ions (Pi) 
will be singlet entangled  - dashed purple line in (d) - even after they
leave the enzyme pocket.
} 
\label{Posner_pair}
\end{figure}

Consider shape deformations of the enzyme pocket
which either increase the distance between the two magnesium ions along the y-axis (decreasing $v_y$) or bring them closer together (increasing $v_y$).
When $v_y$ is greatly increased and the PPi ion is oriented along the y-axis,
weak chemical bonds should form between the end oxygens of PPi and the two magnesium ions,
as depicted in Figure 2c.  By thereby ``pulling" on the electrons of PPi, the magnesium ions
will weaken the chemical bonds between the phosphorus ions and the central oxygen, which will then
be susceptible to hydrolysis
driving the transition, $\text{P}_2\text{O}_7^{4-} + \text{H}_2\text{O} \rightarrow 2 \times \text{HPO}_4^{2-}$.
The liberated phosphate ions (Pi) can then leave the pocket, completing the reaction (see Figure 2d).

The effect of the spins on this reaction can be revealed by retaining 
only four orbital configurations, with PPi aligned along the preferred orientations, $|x_\pm \rangle, |y_\pm \rangle$.
The singlet/triplet wavefunctions have only two states each, $|x_{s/t} \rangle =  ( |x_+ \rangle \pm | x_- \rangle )/\sqrt{2}$, with the same for $x \rightarrow y$.   
The orbital dynamics can be described by a simple Hamiltonian,
\begin{equation}
H_s =- t_s ( |x_s \rangle \langle y_s |  + c.c.) - v_x  |x_s \rangle \langle x_s | - v_y  |y_s \rangle \langle y_s | ,
\end{equation}
with  $s \rightarrow t$ for the triplet sector.
The first term describes rotational motion
while the other terms are the ``pinning" potentials from the magnesium ions. 
Crucially, due to their different symmetries
the singlet and triplet rotation rates are very different $t_{s/t} = t^+_{\text{rot}} \pm t^-_{\text{rot}}$, where $t^+_{\text{rot}},t^-_{\text{rot}}$ are the rates for the ion to rotate
clockwise or counterclockwise, respectively (see Figure 2b).
Crucially, when $t^+_{\text{rot}} = t^-_{\text{rot}}$ as we now assume, the triplet wavefunction cannot rotate, $t_t=0$, due to a destructive quantum interference between the 
clockwise and counterclockwise ``trajectories".

Consider now a shape deformation of the enzyme pocket to drive the chemical reaction, by varying $v_y$ at fixed $v_x$,
starting with $v_y < <v_x$, where both the singlet and triplet ground states will have PPi oriented along the x-axis (Figure 2b).  Now gradually increase $v_y$ until $v_y >> v_x$.
Under this evolution the singlet wavefunction will rotate and align with the y-axis,
where it can bind to the magnesium ions driving the chemical reaction (Figure 2c).
However, the triplet wavefunction cannot rotate, and will get ``stuck" along the x-axis, unable to take part in the chemical reaction.
Strikingly, {\it after the reaction the phosphorus
nuclear spins in the two separated phosphate ions will be entangled in a singlet},
as depicted by the purple dashed line in Figure 2d.
Relaxing the spatial symmetry that gave $t_t=0$ will lead to a non-zero, but small,
probability of triplet entangled phosphate ions being released.

If these (singlet) entangled phosphate pairs are released into the extracellular fluid, they can combine 
with calcium ions to form Posner molecules, where the phosphorus nuclear spins can be ``held" in memory.   Moreover, {\it if two such Posner molecules 
share an entangled phosphate pair, their spins will be entangled}, as depicted in Figure 3a.
Generally, one can envisage complex clusters of highly entangled Posner molecules (see Figure 3b) providing an ideal setting for quantum processing, as we next discuss.

\subsection{Quantum processing with Posner molecules}

\label{sec:Posner-processing}

Consider first the spin and rotational states of a single Posner molecule.
Quantum chemistry calculations find a cubic arrangement for the calcium ions  (eight at the cube corners and one at the cube center)
but the phosphate ions on the six faces reduce the cubic symmetry to $\text{S}_6$, with one 3-fold symmetry axis along a cube diagonal (see Figure 1).\cite{Posner_theory_2000,Posner_theory_2001,Posner_theory_2003,M_Swift} The quantum states of the six phosphorus spins can be labelled by the
total spin, $\text{I}_{\text{tot}} = 0,1,2,3$, and by a ``pseudo-spin",
$\sigma = 0, \pm 1$, encoding the transformation properties under a 3-fold rotation, $| \sigma \rangle \rightarrow \omega^\sigma | \sigma \rangle$
with $\omega = e^{i 2\pi/3}$.
The quantum state with both spin and rotations can be expressed as,
\begin{equation}
| \Psi_\text{Pos} \rangle = \sum_\sigma   c_\sigma | \psi_\sigma \rangle  |\sigma \rangle ,
\end{equation}
with the choice of normalizations $\langle \sigma | \sigma \rangle = \langle \psi_\sigma | \psi_\sigma \rangle =1 $ and $\sum_\sigma |c_\sigma |^2 =1$.
The orbital wavefunction $\psi_\sigma(\varphi)$ depends on the angle $\varphi$ of rotation about the 3-fold symmetry axis.
Fermi statistics requires $| \Psi_\text{pos} \rangle$ be invariant under a 120 degree rotation that interchanges the positions and spins of the phosphorus ions,\cite{E_Bright_Wilson}
implying,  
$\psi_\sigma(\varphi + 2\pi/3) = \bar{\omega}^\sigma \psi_\sigma (\varphi)$
with $\bar{\omega} = \omega^*$.
{\it The nuclear spin and rotational states are thus quantum entanglement
in the Posner molecule}.

The quantum state for two Posner molecules ($a$ and $a^\prime$) can be expressed as,
\begin{equation}
| \Psi_{a a^\prime} \rangle = \sum_{\sigma \sigma^\prime}  C^{aa^\prime}_{\sigma \sigma^\prime} 
| \psi^a_\sigma \rangle | \psi^{a^\prime}_{\sigma^\prime} \rangle  | \sigma \sigma^\prime \rangle_{aa^\prime},
\end{equation}
with (normalized) orbital states depending on the rotation angles $\varphi$ and $\varphi^\prime$,
multiplying a common (normalized) spin state $| \sigma \sigma^\prime \rangle_{aa^\prime}$ which encodes (possible) entanglement between the
spins in the two Posner molecules.   Under a 120 degree rotation about the respective 3-fold symmetry axes, the spin state is multiplied by $\omega^\sigma$ and $\omega^{\sigma^\prime}$, with the orbital states multiplied by compensating factors, $\bar{\omega}^\sigma$ and $ \bar{\omega}^{\sigma^\prime}$. 
The wavefunction $C^{aa^\prime}_{\sigma \sigma^\prime}$, a $3 \times 3$ complex matrix satisfying 
$\sum_{\sigma \sigma^\prime} |C^{aa^\prime}_{\sigma \sigma^\prime} |^2 =1$, encodes {\it pseudo-spin and rotational entanglement between the two Posner molecules}, 
inherited from the spin entanglement provided,
$C^{aa^\prime}_{\sigma  \sigma^\prime}  \ne c^a_\sigma c^{a^\prime}_{\sigma^\prime}$.

Quantum processing with spins in the brain will require ``projective measurements",
induced by chemical reactions which can stimulate subsequent biochemical activity.
The chemical binding of Posner molecules, present in vitro in experiment,\cite{Posner_bind_expt_2013,Posner_bind_expt_2013a} might provide such a mechanism.
Consider two Posner molecules that approach one another
oriented with anti-parallel 3-fold symmetry axis (see Figure 1).
Quantum chemistry calculations reveal that a chemical binding is then possible, and will lower their energy by roughly an electron volt.\cite{M_Swift}
This reaction can be described in terms of the angles of rotation, $\varphi$ and $\varphi^\prime$, about the 3-fold symmetry axis of the two Posner molecules.
Attracted by 
Van der Waals forces the two Posner molecules might stick and (rapidly) rotate on one another,
putting $\varphi = \varphi \equiv \phi$.
But chemical binding will require stopping their rotations.

The dynamics of the angle $\phi$ can be described in terms of a common wavefunction, $\chi_{\sigma \sigma^\prime} ({\phi}) = \psi^a_\sigma(\phi) \psi^{a^\prime}_{\sigma^\prime} (\phi)$, 
which transforms as, $\chi({\phi}+ 2\pi/3) =  \bar{\omega}^{\sigma + \sigma^\prime} \chi(\phi)$, 
and satisfies a Schrodinger equation, $H \chi = E \chi$, with Hamiltonian,
\begin{equation}
\hat{H} =  \hat{\ell}^2/2{\cal I_\text{pair}}  + V(\hat{\phi})  .
\end{equation}
Here $\hat{\ell} = -i \hbar \partial_\phi$ and ${\cal I}_\text{pair}$ is the moment of inertia of the two co-rotating Posner molecules.  Due to the $\text{S}_6$ symmetry, the potential of interaction will satisfy,
$V(\phi + 2\pi/6) = V(\phi)$.
To induce chemical binding we take a very strong (delta-function) interaction,
$V = - V_0 \sum_{n=1}^6 \delta(\phi - 2\pi n/6 )$, and seek bound state solutions with 
$E < 0$.  Such bound states, which correspond to chemically bonded Posner molecules, exist only if 
$\sigma + \sigma^\prime = 0$.  Thus the {\it binding reaction of two Posner molecules induces a ``projective measurement" onto
a state with zero total pseudo-spin}, releasing an electron volt of energy!
The probability that a Posner pair binds (after sticking) is
$P^{aa^\prime}_{\text{react}}  = \sum_{\sigma \sigma^\prime} | C^{aa^\prime}_{\sigma \sigma^\prime} |^2 \delta_{\sigma+\sigma^\prime,0}$.
Once chemically bound the Posner molecules can no longer rapidly rotate,
and are presumably easier to melt via hydrolysis, as discussed in the next section.

\begin{figure*}
\includegraphics[width=7in]{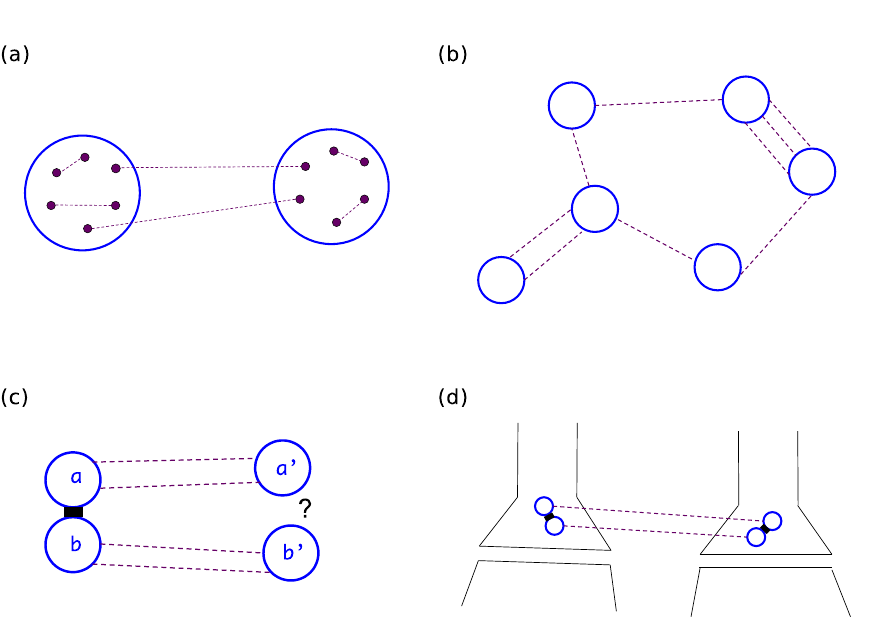}
\caption{
A pair of entangled Posner molecules in (a).  The
purple dashed lines represent singlet entangled phosphorus nuclear spins.  A complex of highly entangled Posner molecules in (b).
With two pairs of entangled Posner molecules, labelled $(a,a^\prime$) and $(b,b^\prime$) as in panel (c), a chemical binding between one member in each pair - the black box connecting $(a,b)$ - 
can change the probability of a subsequent binding of the other members of the pair, $(a^\prime,b^\prime)$.    If the Posner molecules chemically bind after being transported into two presynaptic neurons as depicted in (d), they will be susceptible to melting, 
releasing their calcium into the cytoplasm enhancing
neurotransmitter release, thereby stimulating (quantum) entangled postsynaptic neuron firing.
} 
\label{Posner_pair}
 \end{figure*}

\subsubsection{Quantum Entangled Chemical Reactions}

The chemical binding of multiple Posner molecules with entangled nuclear spins might
allow for complex quantum processing.  
Consider a simple example of 
two entangled pairs,
$|\Psi_{aa^\prime} \rangle \otimes |\Psi_{bb^\prime} \rangle$,
created and situated as in Figure 3c.
We introduce a variable $r =0,1$ with $r=1$ when a reaction binding the Posner pair $\{ ab\}$ proceeds and 
$r=0$ when it does not, and another variable 
$r^\prime =0,1$ for the Posner pair $\{ a^\prime b^\prime \}$.
The joint probability distribution function, $P_{rr^\prime}$, for these two reactions
($P_{11}$ the probability that both reactions proceed, for example)
can be expressed in terms of their common wavefunctions as,
\begin{equation}
P_{rr^\prime} = \sum_{\sigma_a \sigma_{a^\prime}} \sum_{\sigma_b \sigma_{b^\prime}}
|C^{a a^\prime}_{\sigma_a \sigma_{a^\prime}} |^2 |C^{b b^\prime}_{\sigma_b \sigma_{b^\prime}} |^2
g_r(\sigma_a, \sigma_b) g_{r^\prime}(\sigma_{a^\prime} ,\sigma_{b^\prime}) 
\label{Reaction_prob}
\end{equation}
with $g_1 (\sigma , \sigma^\prime) = \delta_{\sigma + \sigma^\prime,0}$ and $g_0  = 1 - g_1$.

Being interested in quantum entanglement between these two reactions, we define
an ``entanglement measure",
${\cal E} =  [ \delta r \delta {r^\prime} ]$,
where $\delta r = r - [r]$ and $\delta r^\prime = r^\prime - [ r^\prime ]$.
The square brackets denote an average with respect to $P_{rr^\prime}$, with
$[ f_{rr^\prime} ] = \sum_{rr^\prime} P_{rr^\prime} f_{rr^\prime}$, for an arbitrary function $f_{rr^\prime}$.  The quantity ${\cal E}$ will depend on the quantum state of the four Posner molecules.

With no entanglement between the four Posner molecules, the wavefunctions take a product form, $C^{aa^\prime}_{\sigma  \sigma^\prime}  =  c^a_\sigma c^{a^\prime}_{\sigma^\prime}$ and  $C^{bb^\prime}_{\sigma  \sigma^\prime}  =  c^b_\sigma c^{b^\prime}_{\sigma^\prime}$, as does the distribution function,
$P_{rr^\prime} = p_r p^\prime_{r^\prime}$.  One can readily verify from 
Eq.(\ref{Reaction_prob}) that this corresponds to 
${\cal E}=0$.  A positive value, ${\cal E} > 0$, implies an enhancement in the tendency for
both reactions to proceed together,
while ${\cal E} < 0 $ reflects an anti-correlation - when one reaction proceeds the other is less likely to, and vice versa.   For generic entanglement between the spins in the Posner molecule pairs, one will have ${\cal E} \ne 0$, indicating that {\it the chemical reactions themselves have become quantum entangled, even if spatially separated}.
Clouds of multiple entangled Posner molecules can induce correlated, non-local 
binding reactions, a powerful setting for quantum processing.

\subsection{Quantum processing with neurons}

To be functionally relevant in the  brain, the dynamics and quantum entanglement of the phosphorus nuclear spins must be capable of modulating the excitability and signaling of neurons - which we take as a working definition of ``quantum cognition".
Phosphate uptake by neurons might provide the critical link.  
In presynaptic glutamatergic neurons the vesicular transmembrane protein VGLUT brings glutamate into the vesicles driven by proton gradients\cite{BNPI_1995a,BNPI_glutamate_1998,VGLUT1_2000} (the vesicle is acidic\cite{Vesicle_pH_2004}
with pH = 5.5).
In the original discovery paper\cite{BNPI_1994} in 1994, 
VGLUT was reported to have a sequence homology to a (rabbit) kidney phosphate transporter,
which brings phosphate into cells 
driven by a sodium concentration gradient.\cite{Na/Pi_1978, Na/Pi_transporter_2012}
Moreover, VGLUT (which was, at the time,
named BNPI for brains sodium-phosphate transporter\cite{BNPI_1994})
was found to uptake phosphate when 
expressed in Xenopus oocytes, in a sodium-concentration dependent manner.
We propose that VGLUT plays a {\it dual physiological role}, both transporting glutamate into
presynaptic vesicles and 
transporting phosphate ions into the presynaptic neurons  during vesicle endocytosis\cite{Kandel_Neuro}  - and that this enables neuron uptake of Posner molecules, as detailed below.

A rapid influx of calcium following an incoming action potential triggers the presynaptic vesicles
to fuse with the cell wall and release 
glutamate into the synaptic cleft (exocytosis).\cite{Kandel_Neuro}  
During subsequent endocytosis these vesicles are retrieved (from the cell wall or in a ``kiss-and-run" mode\cite{Vesicle_recyling_2003}) and brought back into the presynaptic neuron.
In this process the sodium-rich extracellular fluid (with pH=7.4) will enter the vesicle,
perhaps engulfing Posner molecules floating in the synaptic cleft.
After pinch off and retreat\cite{Kandel_Neuro} the vesicle interior will become acidic due to  
proton pumps.
Once the pH drops below 6, we anticipate
that any enveloped Posner molecules will melt via hydrolysis (``proton attack")
releasing phosphate and calcium ions into the vesicle interior.

Due to the high $\text{Na}^+$ concentration in the vesicle interior after endocytosis, the transmembrane protein VGLUT, now exposed to a large sodium concentration gradient,\cite{Na/Pi_1978} 
might transport the phosphate ions out of the vesicle into the cytoplasm.
With local cytoplasmic calcium levels elevated during exocytosis,
these phosphate ions could recombine with calcium, forming Posner molecules inside the neuron.
In effect, {\it glutamate release has triggered the influx of Posner molecules 
into the presynaptic neurons}.

If a chemical bond subsequently forms between two Posner molecules in the lower pH=7 environment of the cytoplasm, the 
stationary (non-rotating) dimer will be susceptible to melting via hydrolysis (``proton attack") - liberating  18 calcium ions which could  stimulate further glutamate release, thereby enhancing the firing of the postsynaptic neuron.

During cellular uptake, nuclear spin entanglement between two different Posner molecules 
will be retained, even if transported into two different neurons.
The uptake of many Posner molecules could induce nuclear spin entanglement between multiple presynaptic neurons.
The chemical binding and subsequent melting of two Posner molecules inside a given neuron would then
influence the probability of Posner molecules binding and melting in other neurons.
{\it This could lead to non-local quantum correlations in the glutamate release and postsynaptic firing across multiple neurons}. 
 
A simple example with two neurons illustrating this critical link between nuclear spin entanglement and neuron firing rates is depicted in Figure 3d.
Compound and more elaborate processes involving multiple Posner molecules and multiple neurons are possible, and 
might enable complex 
nuclear-spin quantum processing in the brain.

\subsection{Prospects}

In this paper an apparently unique mechanism for quantum processing in the brain
has been explored.  The phosphorus nuclear spins in phosphate ions serve as qubits, pairwise entangled 
during hydrolysis of pyrophosphate, engulfed and protected inside Posner molecules,
inducing entanglement of the nuclear spins and rotational states of multiple Posner molecules,
which can be transported into presynaptic glutamatergic neurons during vesicle endocytosis, with intra-cellular calcium being released by subsequent binding and melting of the Posner molecules, stimulating further glutamate release, thereby enhancing,
and quantum-entangling, postsynaptic neuron excitability and activity!

An intricate story, with multiple links in the chain of required processes.
We briefly mention some experiments that might serve to refute, or perhaps strengthen, the hypothesis of nuclear spin quantum processing in the brain.

Dynamic light scattering and cryoTEM could be employed to explore
the concentration of Posner molecules in simulated body fluids,
upon varying the ion concentrations, pH and other control variables.\cite{Posner_expt_1998,Posner_expt_1999,Posner_expt_2010}
(Attempting to establish whether Posner molecules are present in real body fluids,
while challenging, would also be critical.)   Liquid state NMR methods could be used to 
measure the spin dynamics (e.g. spin coherence times) of the phosphorus nuclei inside Posner molecules.\cite{NMR_Hore}  Calcium and oxygen isotopes (with non-zero nuclear spin) if incorporated
into the Posner molecules would presumably decohere the phosphorus nuclear spins,
which might be accessible with NMR.
Determining the prospects of replacing the central calcium ions in the Posner molecule with  ``impurity" elements - for example lithium and mercury ions, energetically favorable in quantum chemistry calculations\cite{M_Swift} - would also be instructive.   If replacement is possible, varying the lithium and mercury isotopes
and examining the effects on phosphorus spin coherence inside the Posner molecules could also be interesting.

Many aspects of the mechanisms proposed in this paper could be explored in vitro.
Establishing control of pyrophosphate hydrolysis 
catalyzed by the enzyme pyrophosphatase,\cite{Kandel_Neuro,Alberts_Cell,Lodish_Cell} in vitro would be a first step.
With calcium present the released phosphate ions should bind into Posner molecules.
Probing possible phosphorus nuclear spin entanglement between multiple Posner molecules
might be possible by separating the solution into two separate containers,
lowering their pH to induce melting of chemically bonded Posner molecules and measuring the 
calcium release with calcium fluorescence molecules.  
Quantum entanglement would be revealed by coincidences and correlations
between the fluorescence emitted from the two containers.
If present, one could envisage performing quantum processing,
and, conceivably, designing  and implementing a liquid state nuclear-spin quantum computer.\cite{Qu_Computing}

The mechanism for neuron uptake of Posner molecules, arguably required
for in vivo quantum processing with phosphorus nuclear spins, relies
on the transport of phosphate by VGLUT from the presynaptic vesicle interior into the cytoplasm.\cite{BNPI_1994}
In vitro experiments further establishing and characterizing the potential (sodium-concentration driven) phosphate transport 
by VGLUT would be essential.\cite{BNPI_1995}

If the phosphorus nuclear spins inside Posner molecules are playing a functional
role in the brains of mammals (or, possibly, other vertebrates), then perturbations of the nuclear spins might have behavioral manifestations.   Strong time and spatially dependent magnetic fields would be expected to modify the phosphorus spin dynamics inside Posner molecules,
and could be characterized with NMR.
Might this inform trans-cranial magneto stimulation protocols,\cite{TMS}
modifying their efficacy in treating mental illness?
If two lithium ions can be incorporated inside the Posner molecules during molecule formation
(replacing the central divalent calcium cation) they would tend to decohere
the phosphorus nuclear spins, offering a possible explanation for the remarkable efficacy of lithium in tempering mania in patients with bipolar disorder. 
If this is indeed the mechanism, 
one might expect a lithium isotope dependence on the behavioral response.
Remarkably, a lithium isotope dependence on the mothering behavior of rats chronically fed either $^6\text{Li}$ or $^7 \text{Li}$ - having elevated or depressed alertness levels, respectively - has indeed been reported.\cite{Li_Rat_1986}   Reproducing this striking experiment would be paramount.
Chronic ingestion of the calcium-43 isotope, which has a large $\text{I}=7/2$ nuclear spin,
might also possibly have deleterious effects on mice and rats.
Might an exploration of the effects of
shock waves on the mechanical stability and nuclear spin dynamics 
(induced via excitation of vibrational modes) of Posner molecules free floating in water have some relevance to brain trauma?\cite{Brain_Trauma}

It is hoped that
the various experiments suggested above might be informative, in and of themselves - and possibly in refuting, or supporting, the hypothesis of nuclear-spin quantum processing in the brain.

\acknowledgements
I am grateful to many, many physicists, chemists and neuroscientists for invaluable
input and suggestions, including,
Ehud Altman, Leon Balents, Bill Bialek, David Cory, Robert Edwards, Marla Feller, Daniel Fisher, Michael Fisher, Yvette Fisher, Craig Garner, James Garrison, Mike Gazzaniga, Steve Girvin, Tarun Grover, Songi Han, Paul Hansma, Matt Helgeson, Alexej Jerschow, Ilia Kaminker, Charlie Kane,
David Kleinfeld, Ken Kosik, Allan MacDonald, Michael Miller, Thomas Mueggler, Ryan Mishmash, Lesik Motrunich, Nick Read, Richard Reimer, Jeff Reimer, Steve Shenker, Boris Shraiman, T. Senthil, Chuck Stevens, Ashvin Vishwanath, Shamon Walker, David Weld, Xiao-Gang Wen and Xuemei Zhang.
I would like to especially acknowledge Michael Swift for his quantum chemistry calculations of Posner molecules. 
This research was supported in part by the National Science Foundation under Grant No. DMR-14-04230, and by the Caltech Institute of Quantum Information and Matter, an NSF Physics Frontiers Center with support of the Gordon and Betty Moore Foundation (M.P.A.F.). 
Thanks to the Aspen Center of Physics where some parts of this work were completed.

\bibliographystyle{mysty}
\bibliography{Qu_neuro-synopsis}
\end{document}